# Validation of Hardware Security and Trust: A Survey

Payman Behnam
Department of Computer Science, School of Computing
University of Utah, USA
payman.behnam@utah.edu

*Abstract*— With ever advancing in digital system, security has been emerged as a major concern. Many researchers all around the world come up with solutions to address various challenges that are crucial for industry and market. The aim of this survey is a brief review of challenges of security validation as well as define and classify Hardware Trojans. Then, we provide more details about various validation techniques for hardware security and trust.

## I. INTRODUCTION

Secure designing and removing vulnerabilities of a design are becoming a major concern in the semiconductor industry. We are observing the growing computing platforms in many areas such as smart cards, financial systems and many Internet of Thing (IoT) devices. These are high-risk applications of computing that need a lot of attentions to enhance their secure platforms. Due to following reasons the demand for providing security solutions has been drastically increased and virtually all companies are trying to meliorate security assurance [42].

*Increasing skills and resources of adversaries:* Nowadays, we can observe many attacks that have been launched for economic or political reasons with novel approaches. For example, Stuxnet worm cause the leak of sensitive information from a specific nation state [45].

*Increasing attacks in hardware-oriented:* Previously, adversaries attack a system by targeting only software components in a way that get privileges for executing a malevolent code [43]. In the recent years, there is large trends toward hardware security with the aim of opening backdoor to simplify bypassing of protection mechanisms by operating systems [44]. Furthermore, detection of malicious circuit is very hard and their harms to circuit are considerable [42]. Besides, by concerning time to market, System-on-chip (SoC) designers make use of Intellectual Properties (IPs) to expedite the production process. However, these IPs are from all over the world and hence they are not trusted. Therefore, security validation is an essential need for the IPs itself.

Hardware Trojans (HTs) are malicious circuits that are added to design by adversaries to extract the valuable information or defeat the trustworthiness of design or modify the functionality of the design. Figure 1 shows electronic supply chain security model. As you can observe, HT can be added in almost all stages including, specification, IP integration, implementation, synthesis, physical design, manufacturing, etc. However, the attackers tend to add HTs in the design stage or IP integration rather than in fabrication phase since they don't need to access foundry facilities to insert HTs [51].

In this survey, we are going to study the various challenges and solution of trust validation. The remaining part of the article is arranged as follows. Section II expresses validation importance in VLSI design cycle. Section III is about security validation challenges with functional design approaches. Section IV explains different classes of hardware Trojans. Section V is the main part of the paper and reviews verification techniques for hardware trust whereas Section VI concludes the paper.

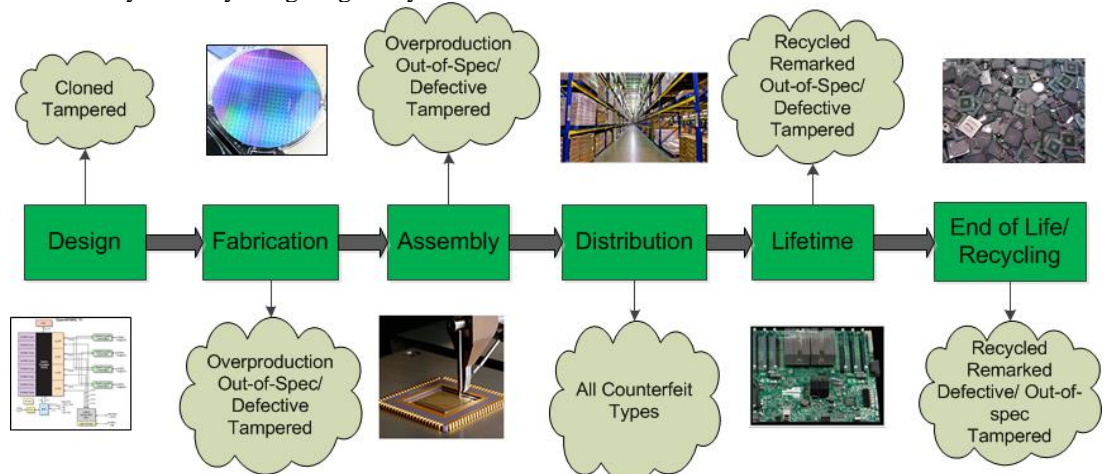

Figure 1-Electronic supply chain security model [47].

## II. VALIDATION IMPOTANCE

Validation means to be sure that the design implementation satisfies design specification requirements. Diagnosis and correction mean scrutinizing an implementation to find errors and remove them while it is correspondent to the specification respectively [1]. With the ever-increasing of digital system complexity, the specific properties of the design are growing and some corner cases and unreachable states are becoming more than ever. Hence verification and validation of systems is an important and challenging task while about 60% of design effort is dedicated to design verification [1][2].

The validation approach can be categorized into formal (static) [3], semi-formal [4] and simulation (dynamic) methods [5].

The formal methods are based on mathematical features and can be classified in theorem proving, equivalence checking and property (model) checking. The advantage of using this method is high coverage. Thus, one can assure that all possible situations have been verified. Nevertheless, they are complex, and generally non-user friendly. Besides, it is difficult to provide mathematical model for all designs especially for large one [3].

The simulation mechanisms are based on input-output data and comparing the output results with expected ones. These approaches are easy to use, have low cost and can be used for verification of large designs. However, their error coverage is low. Hence, some errors may be skipped in verification phase [4].

In the semi-formal approaches, it is tried to make use of the advantage of both formal and simulation-based methods. They can be categorized in assertion-, coverage- and symbolic simulation-based methods [1].

In the last decades, several new mechanisms such as modern decision diagrams [5][6][7], abstraction [8], satisfiability at binary level [9][10][11], satisfiability modulo theory [12], and bounded model checking [13] have been emerged. There are also some trends in using mutation [14], symbolic algebra [15], and probabilistic approaches [16] in verification, debugging and correction phases.

## III. SECURITY VALIDATION CHALLENGES

In Security validation like design validation, the goal is investigating some specific properties of a system (e.g. in this case the security properties). However, there are several challenges that cause we cannot make use of design validation techniques directly [17].

The first challenge is that vulnerabilities or Trojans may hardly change the functionality of systems. Therefore, direct using of functional verification technique is not useful at all [18]. The second issue is that some Trojan can be activated because of component interactions. Accordingly, unlike design verification, security validation may be postponed until all components are integrated together. The third issue is that defining properties for security issues is not straightforward. Converting security statements into exact requirements needs expertise in design verification and security domain [19]. The final point is that in the security validation we do not have specific model and well-defined criteria. Consequently, it is not possible to decide that existing test cases [50] are enough or not [17]. Moreover, there are some trade-off between debugging and security issues. While more observability is advantageous for debugging purpose, it is not suitable for security.

To overcome these problems, several solutions have been proposed. One of the efficient way for security amelioration is to begin the execution of Secure Development Lifecycle (SDL) as early as possible. In this manner, one can solve resource conflicts that effect test environment implementation and improve threat models. Pay attention that with limited resources one cannot cover all test cases. Therefore, some changes that effect on device security objectives can be identified [17].

One of the efficient way for utilizing functional validation for security trust and validation is fuzzing or random input testing methods [20]. Fuzzing method is to some extent similar to constraint testing. However, there are few discrepancies between fuzzing and constrained random testing. Fuzzing testing methods try to apply unexpected inputs in various methods, while constrained testing strategies may overlook such inputs altogether. It has been proved that by combining protocol knowledge, mutation and randomness, fuzzing approach is an efficient way in discovering vulnerabilities in both software and hardware [20]. Besides, they are easy to implement with a little overhead. By weighting the valid and invalid inputs and adjusting them, one can enhance coverage and efficiency of them in detecting vulnerabilities [17].

Static analysis approaches are recommended in design validation methods. Compromised firmware that is patched to systems imposes important security risks. In this situation, we can make use of static analysis tools for all firmware with access to system resources to reduce the risk of security issues [21].

## IV. HARDWARE TROJAN CLASSIFICATION

Hardware Trojans are composed of trigger (activation mechanism) and payload (malicious function). The input of trigger functions is hardly activated; so, it is difficult for test cases to detect hardware Trojans. In one perspective, Trojans can be classified in two categories: The first type is combinational Trojans and the second one is sequential ones. If the Trojan is activated based on specific condition or internal signals or flip-flops, it is called combinational Trojans. However, in sequential ones, the Trojans are stimulated based on specific FSM that are activated when a sequence is provided. You can see the differences in Figure 2.

In another perspective, Hardware Trojans can be grouped based on their effect on the normal functionality of a design. If it directly changes the normal outputs, it is called bug-based HT. However, if it does not have any impact on the normal functionality, but add some extra logic along with original design it is called parasite-based HT [22].

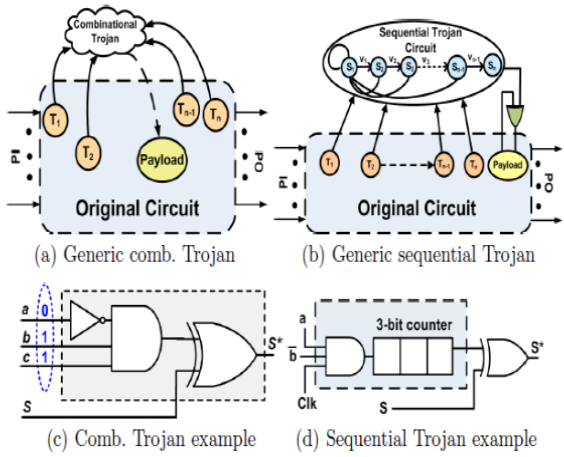

Figure 2-Different Types of Trojans [29].

Figure 3 and 4 show HT classification in both circuit structure and K-Map structure. Figure 3(a) depicts the normal circuit whereas Figure 3(b) sketches the circuit while an attacker may convert the original circuit to a malicious one by adding an inverter on the $d_1$ as an input. In this case $d_1$ can serves as original and trigger input. In general, they are not suitable for attacker since they can be detected easily in design verification stage. Hence, almost all HTs in designs are parasite-based type. Suppose that an attacker wants to add the former malicious circuit $f_m$ into design while the original one is $f_n$. To do so, he can add the trigger function $t_1 t_2$ and control when the malicious circuit is going to be activated Figure 3(c). Figure 4(a) to Figure 4(d) show the corresponding K-Map of circuits in Figure 3(a) [22].

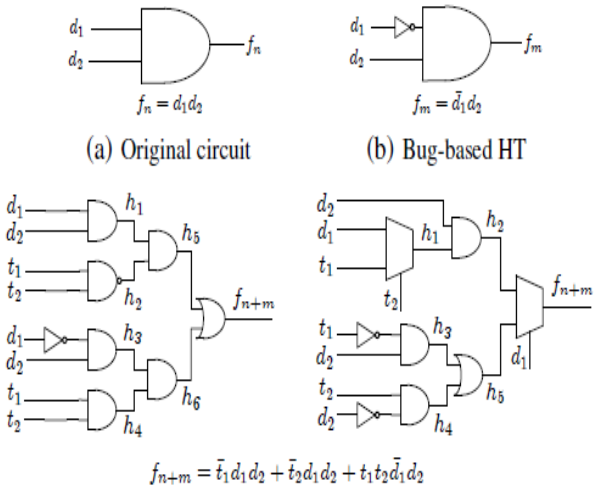

Figure 3- HT classification in both circuit. (a) original circuit, (b) Bug based type, (c) parasite based type while trigger condition is {$t_1$, $t_2$} = {1,1} [37]

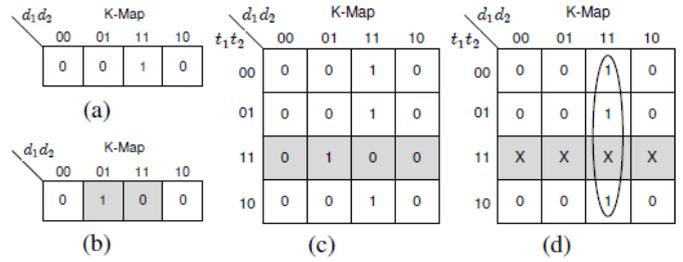

Figure 4- K-Map Model od circuits in Figure 3. (a) for original circuit, (b) for Bug-based type, (c) for parasite-based type, (d) setting the third row of (c) as don't care [37]

## V. VERIFICATION TECHNIQUES FOR HARWARE TRSUT

Generally, there are there categories for validation of hardware trust. They are functional, formal and trust verification approaches [51, 22]. The third one is a new category which has different names in the literature and it is taken from the work in [22]. Each of these categories has several sub-categories. Figure 5 shows these categories in detail. In the following subsections, we review each of them in more details.

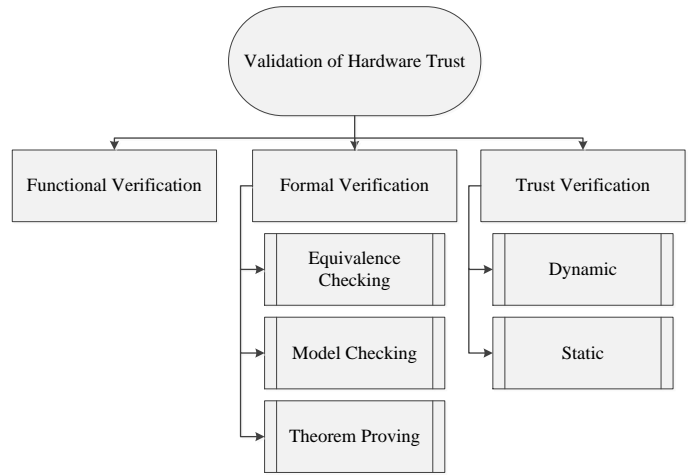

Figure 5- Classification of hardware trust validations [22]

### A. Functional (Simulation-based) Verifcation

Hardware Trojan in functional (Simulation-based) verification can be distinguished by galvanizing and monitoring in the functional verification stage. However, as mentioned in Section IV, this method works for bug based hardware Trojan. For parasite type, this method rarely works if not possible [22]. Automatic Test Pattern Generation (ATPG) and Code Coverage can be utilized for distinguishing circuit with small Trojan and circuits without them. The effective test patterns for finding parasite trojans are hard to generate. The authors in [28] suggested a method to localize HT in four steps. First, The ATPG is used for catching easy-to-detect signals. Afterwards, a full scan N-detect ATPG is used to identify hardly activated nodes and finding suspicious circuits. At that point, a SAT solver is exploited to diminish the search space of mistrustful nodes. In the last step, the gates that are not covered are grouped by region separation approach to discover HTs. This method needs a golden netlist and due to using SAT mechanism is not scalable enough [51].

A statistical approach technique called MERO in [29] engender test patterns to trigger rare nodes several times (Figure 6). In this manner, they can enhance the chance of activation. Accordingly, the Trojan detection chance can be increased. Nevertheless, they don't consider payload role and hence the effect of trigger can be masked. To solve this problem, the authors in [30] suggest a solution that exploit genetic algorithm to create test cases that propagate the impact of HT to primary outputs. Then, an ATPG is employed to discover activation patterns of the dormant hardware Trojans. Finally, a code coverage analysis over IPs is performed to detect corner cases in design that can be used as backdoor for adversaries and attackers. Though, HT may still remain in design while we have 100% code coverage. The authors in [31] make use of code coverage along with formal verification. If they fail to distinguish HT, other rules are exploited to discover unused and redundant circuits.

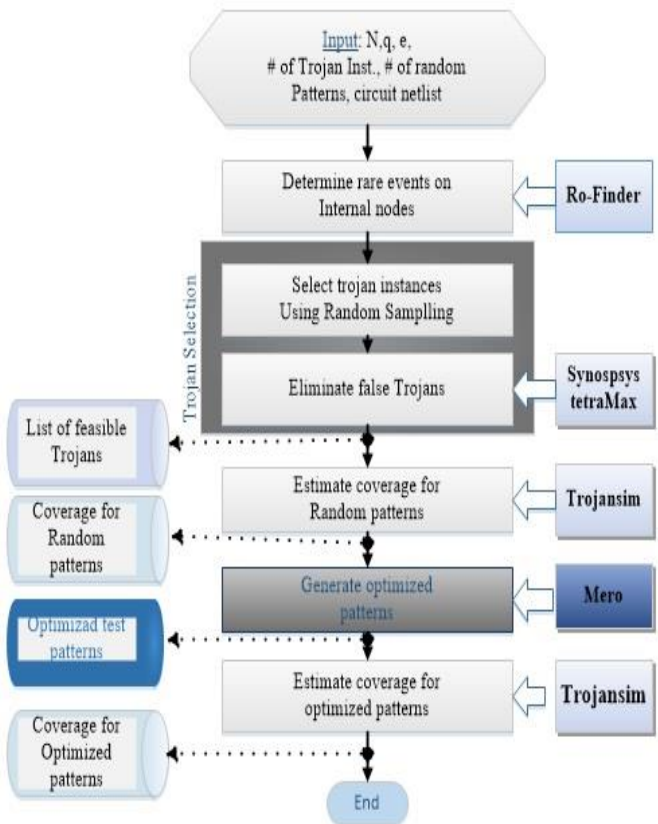

Figure 6- Overview of MERO Approach [29].

### B. Formal verification

The limitations of functional testing methods in coverage has led to emerging formal methods [23]. In some work such as [24, 25, 27] the advantages of formal approaches over testing methods in exhaustive security have been demonstrated carefully. In the following subsections, we review main categories of formal verification for security purpose.

### B.1- Trojan Detection using Equivalence Checking

Equivalence checking is one of the approaches in formal verification [46, 49]. By using this approach, one can make sure two aspects of a design (i.e. specification and implementation) behave similarly.

Figure 7 illustrates a big picture for performing equivalence checking by making use of equivalence checker (e.g. SAT solver). The outputs of two different designs are XORed, then the output of XORs are Ored. Then Conjunctive Normal Form clauses (CNF) of the whole design are passed to a SAT solver. If the implementation and the specification are same, the final output is zero, otherwise one.

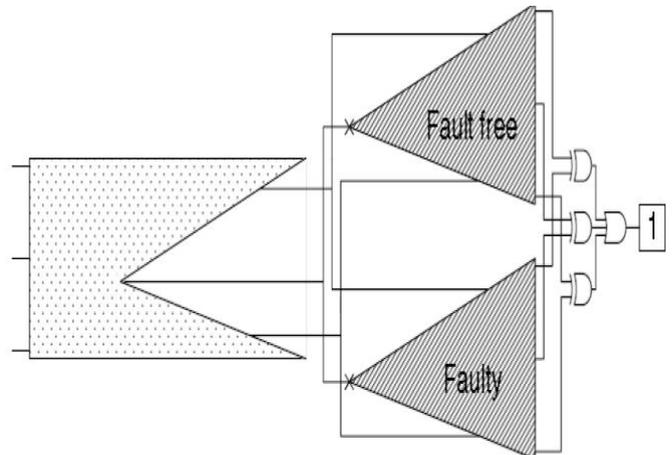

Figure 7- Equivalence checking Mechanism [5]

### B.2- Trojan Detection using Theorem Proving

Theorem proving is one of the powerful but non-user friendly approaches in formal verification. Many theorem provers have been introduced for hardware and software issues from 1960 up to now [23]. One of recent theorem provers is Proof-Carrying Hardware platform (PCH) that is developed for trust validation of soft IPs [33-34]. The core of this solution is based on Proof-Carrying Code (PCC) [32]. By making use of the PCC approach, software developers can certify their software Code. Then, developers can provide PCC binary file with a formal approach. The run time of this method is short and it has compatibility with different applications. Authors in [33–35] proposed the PCH platform for dynamically reconfigurable runtime combinational equivalence checking. This method can be used for RTL and netlist validation. Figure 8 shows the details of this procedure.

### C. Trust Verification

Untrusted circuit can be recognized by trust verification. The approach is established using the fact that HTs are hardly activated. This identification can be performed by detecting unsensitized paths within verification stage in dynamic manner such as UCI [36] and VeriTrust [37] or Boolean functional analysis statically such as FANCI [38].

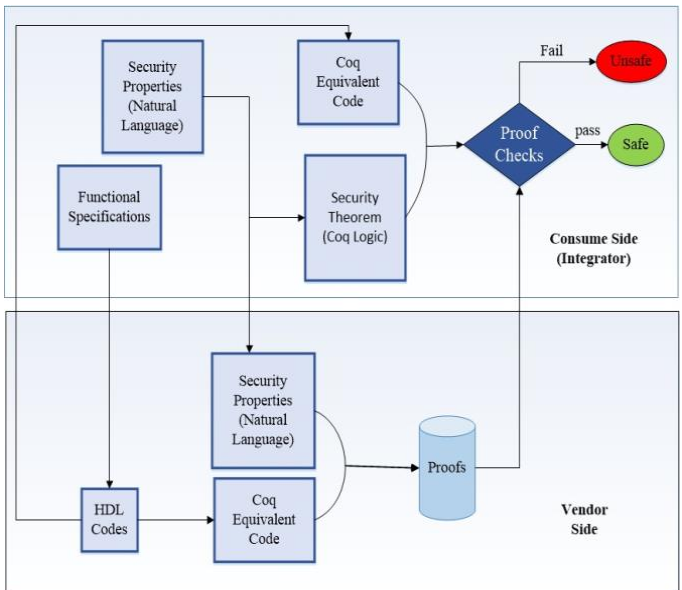

Figure 8-Overview of PCH platform [34]

*D.1 Unused Circuit Identification (UCI)*

The author of UCI proposed a defensive strategy called BlueChip that works design-time as well as runtime. During the design validation, they identify unused circuit that are potential locations for malicious circuits. The UCI builds a flow graph from the circuit and defines the direct and indirect signal pairs. This pairs are equal when there are not any trigger situations [39]. Afterwards, it verifies the circuit to locate unemployed parts (e.g. parts of the design that do not effect on the primary outputs). BlueChip eliminates these circuits and substitutes them. Figure 9 demonstrate a big picture of BlueChip architecture.

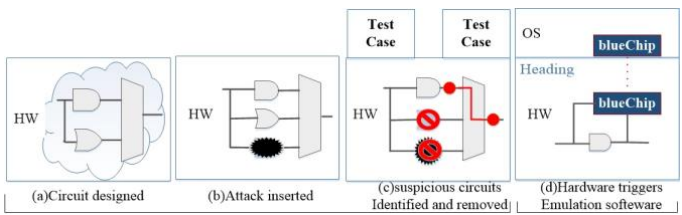

Figure 9- Overview of BlueChip mechanism [36]

*D.2 VeriTrust*

Figure 10 shows the big picture of VeriTrust work [37]. The tracer tracks verification test cases to recognize those signals that comprise non-activated entries of design functionality (i.e. entries of K-Map). The desired signals for tracking are outputs and the inputs of flip-flops.

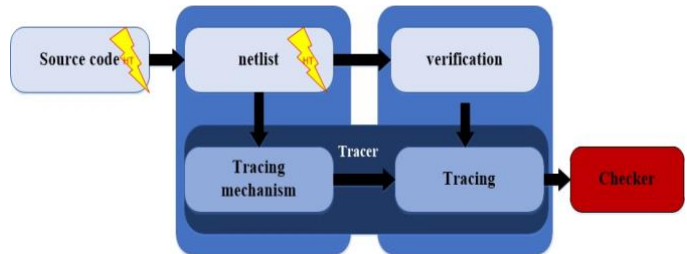

Figure 10- Big picture of VeriTrust Solution

As you can observe in the third row in K-Map in Figure 4 (c). shows the malicious function. When we compare the K-Map of the clean circuit and Trojan inserted circuit we can figure out that the hardware Trojans increases the size of K-Map when the number of inputs is increased. Accordingly, the circuit can exhibit the normal and malicious behavior that are controlled by trigger inputs. Uncovered entries are set as don't care values to detect excessive inputs and hence potential location of hardware Trojans (Fig. 4(d)). By following digital logic rules, one can get the original cinputs ($d_1d_2$). Consequently, the trigger inputs, $t_1$ and $t_2$, are redundant.

The main advantage of VeriTrust is that VeriTrust is not sensitive to the way of implementation of hardware Trojans. Nevertheless, DeTrust [40] introduces a new Trojan type in a way gates are governed using a subset of primary inputs and trigger conditions occurs across several clock cycles that VeriTrust cannot detect them.

*D.3 FANCI*

The author of FANCI [38] propose a novel solution to recognize the unused circuits using their Boolean functions and their values. They make use of the assumption that unused circuits are suitable location for inserting HTs. They make an approximate truth table for internal signals to figure out their impact on outputs. They exploit a predefined threshold for trigger conditions. All signals with lower stimulation than the computed threshold are mistrustful parts. They recommend multiple heuristics algorithms to define backdoors from control value vectors. A main feature of FANCI is that it does not rest on validation result. Furthermore, FANCI is not stalled by noncomplete tests and accordingly works without false negatives [22]. The disadvantage of this method is that several gates in design can be announced as safe gates inaccurately (false positives). For example, In DeTrust [40] the probability of the activation new Trojan types is higher than threshold value and hence FANCI considers them as non-malicious gates which is not true. On the other hand, if the circuit is trustworthy, there is chance that some gates are defined as suspicious functions in accurately.

Table 1-Characteristics of functional and Trust Verification for Hardware Trojan Detection [22]

| Method | Static/Dynamic | Detection Method | Runtime | False Negative | False Positive |
|---|---|---|---|---|---|
| Functional Verification | Dynamic | Activate the HT | Good | HTs with rare trigger condition | None |
| UCI [36] | Dynamic | Identify equal signal pairs | Fair | HTs in [48] | Some with through verification |
| VeriTrust[37] | Dynamic | Identify HT trigger inputs | Fair | Unknown | Some with through verification |
| Fanci[38] | Static | Identify weakly affecting inputs | Fair | Possible with low threshold | Many with high threshold |

### D. Discussion

Table 1 reviews and compares the characteristics of the functional and trust verification. The UCI [36] and VeriTrsut [37] are complement solutions for dynamic trust validation. With growing quantity of test vectors, the possibility of hardware Trojan activation is increased. Accordingly, the quantity of potential untrusted designs detected by these two approaches is decreased. Since the FANCI [38] is a static solution, the attacker could examine their Trojans by exploiting FANCI approach without any hypothesis of the unknown test vectors that in practical is used to detect Hardware Trojans [22].

All trust validation methods try to minimize false positive and remove false negative as much as possible. However, the amount of detection can be affected by input and user specific parameters and inputs. For example, if the number of test cases is reduced, UCI and VeriTrust report many Hardware Trojan signals which is not necessarily true (false positive). Besides, FANCI make use of a threshold value for Trojan detection. If the value of threshold is small, there is chance that some HT-related wires are missed. However, if it is large, it is likely that it detects many safe gates as suspicious ones.

### VI. CONCLUSION

Defining and mitigating security vulnerabilities and Trojans in today's complex designs are a complex tasks and important concerns. Current state of the art procedures have not addressed all the existing challenges. Hence, adversaries exploit the backdoors to steal valuable information or deviate the function of designs. This survey presents the concept and different classes of Hardware Trojans. It also presents the challenges of using design validation technique for security purposes. Then, it reviews different strategies for security validation including functional, formal and trust validation. Especially, it expresses and compares three major works in trust verification including UCI, VeriTrust and FANCI. Although the current valuable approaches have solved some challenge of security validation, there are a lot of problems which need huge efforts to solve the security concerns in various aspects.